\documentclass[letterpaper,final,notitlepage]{iopart}
\usepackage{iopams}
\usepackage{graphicx}
\usepackage{color}
\usepackage{epsfig}
\usepackage{times}
\usepackage{cite}

\renewcommand{\vec}[1]{\mathbf{#1}}

\begin{document}

\title{Transverse confinement in stochastic cooling of trapped atoms}

\author{D. Ivanov\footnote[1]{Also at: Russian Center of Laser Physics,
  St.-Petersburg State University, 5 Ulianovskaya, Pedrodvoretz, St.
  Petersburg, Russia} and S. Wallentowitz\footnote[7]{email:
    sascha.wallentowitz@physik.uni-rostock.de}} 

\address{Emmy--Noether Nachwuchsgruppe ``Kollektive Quantenmessung und
  R\"uckkopplung an Atomen und Molek\"ulen'', Fachbereich Physik,
  Universit\"at Rostock, Universit\"atsplatz 3, D-18051 Rostock, Germany}

\begin{abstract}
Stochastic cooling of trapped atoms is considered for a laser-beam
configuration with beam waists equal or smaller than the extent of the atomic
cloud. It is shown, that various effects appear due to this transverse
confinement, among them heating of transverse kinetic energy. Analytical
results of the cooling in dependence on size and location of the laser beam
are presented for the case of a non-degenerate vapour.  
\end{abstract}


\section{Introduction}

Cooling techniques for atoms play a crucial role in modern physics. They have
allowed for the localisation of atomic gases by weak trapping forces and step
by step have enabled the reach into the domain of ultracold gases. At such low
temperatures the peculiar quantum-statistical properties of the atoms become
immanent in various effects of cold atomic collisions. Bose-Einstein
condensation \cite{bec,bec2,bec3,bec4,bec5} and the recent production of
Fermi-Dirac degenerate gases \cite{fermi,fermi2,fermi3,fermi4} are limiting
cases of the now existing experimental feasibilities. Moreover, the
implementation of atom-lasers \cite{atom-laser,atom-laser2,atom-laser3} and
microstructured traps on so called atom chips
\cite{bec-micro,bec-micro2,bec-micro3,bec-micro4} show the vast potential of
applications.

The typical strategy to generate a Bose-Einstein condensate from a moderately
cold sample of trapped atoms is to apply different cooling techniques in
sequence: it usually starts with laser cooling
\cite{laser-cooling,laser-cooling2,laser-cooling3} and ends with evaporative
cooling \cite{evap-cooling,evap-cooling2,evap-cooling3}. The latter technique
seems to be unbeaten as of yet for the final cooling step. Laser cooling, that
relies on cycling transitions where photons are spontaneously emitted, does
not provide the ultimate cooling power, due to the reabsorption and scattering
of the emitted photons. However, the drawback of evaporative cooling is well
known to be its intrinsic loss of atoms. Since hot atoms are released from the
trapping potential to reach a colder sample, a substantial atom loss has to be
taken into account that ultimately limits the size of the condensed sample.
Furthermore, as does sympathetic cooling \cite{symp-cooling}, it requires
sufficiently strong atomic collisions for thermal re-equilibration.

Given a prepared sample of condensed atoms, a multitude of technical and
possibly fundamental noise effects lead to a finite lifetime of the condensate
state. Among these noise effects there are collisions with background vapour,
electromagnetic noise sources via the trapping potential, scattering of light,
etc. The study of these detrimental disturbances and the development of
methods to reduce their impact on the condensate will be a challenging task
for the future. One way to compensate for such heating effects may be simply
the continuous application of cooling during the entire experiment. This may
partially compensate the heating and thus extends the lifetime of the
condensate. The only technique working so far at these temperatures is
evaporative cooling. Its continuous application, however,
would be rather unfortunate for the condensate, since though the lifetime of
the condensate may be extended, its size in terms of atom number will
continuously decrease.

Some years ago, Raizen et al. have proposed the use of stochastic cooling for
trapped atoms \cite{stochastic-raizen}. It is a successful method in
high-energy physics \cite{stochastic-cooling,stochastic-cooling2} where the
transverse motion of a particle beam has to be collimated and cooled. Clearly,
the energies involved there are not in the regime important for an application
to trapped atoms. However, classical numerical simulations have shown the
feasibility of stochastic cooling also for trapped atoms
\cite{stochastic-raizen,stochastic-raizen-josa}. Furthermore, it has been
recently shown, that also at ultralow temperatures this technique reveals
cooling \cite{stochastic}. Thus it may perhaps be utilized to stabilise an
atomic Bose-Einstein condensate.

It should be pointed out that on the single-atom level feedback control of
atomic position has been theoretically studied \cite{single-fb,single-fb2} and
experimentally realised in optical lattices \cite{morrow} and high-quality
cavity fields \cite{rempe}.

In this paper we extend our analysis of stochastic cooling of trapped atoms to
include also effects due to the transverse confinement of atoms. The latter
has its origin in the finite beam waist of the employed control-laser beam. At
temperatures above the condensation point we give analytic results of the
cooling and discuss its optimisation with respect to size and location of the
control-laser beam waist.
   
In Sec.~\ref{sec:stoch-cooling} we explain the method of stochastic cooling of
atoms and derive the expression for the single-atom density matrix after the
single cooling step. Given this result at hand, in
Sec.~\ref{sec:energy-change} we calculate the total energy change of atoms due
to a single step of stochastic cooling in terms of quantum-statistical
averages. In Sec.~\ref{sec:non-degenerate-atomic-vapor} the regime of a
non-degenerate gas is considered for which analytical expressions for the
energy contributions are obtained. Moreover, the dependence of cooling on
geometrical parameters is discussed. Finally, in Sec.~\ref{sec:conclusions}
conclusions are given.

\section{Stochastic cooling}
\label{sec:stoch-cooling}

The method of stochastic cooling of trapped atoms consists of the repeated
application of two operations: the measurement of the momentum of atoms and
the subsequent application of a kick to compensate for the measured momentum.
Several aspects are important and should be emphasised for an understanding of
the working of this technique. First of all it is not done on a single atom
but on a large set of atoms. Those atoms that are subject to measurement and
kick are specified by their spatial location in a given volume of space. In
the experiment that volume is defined by the spatial extent of the laser
beams that implement the required operations. 

Since in the experiment, at the time of measurement of momentum, it is usually
unknown how many atoms contributed to the measured signal, the momentum per
atom averaged over the atomic ensemble is not the measured observable. To
obtain this momentum per atom one would in fact need knowledge on the precise
number of atoms, that contributed to the measured signal. What can, instead,
be assessed by the measurement is the total momentum of the atoms. This is the
sum over the atomic momenta, since each atom equally contributes to the
signal.

Given the measured total momentum of the set of atoms, for compensating it, a
(optical) field is turned on to provide the necessary kick by its interaction
with the atoms. Since each atom separately interacts with the field, one can
only apply a common kick to each atom. The determination of the required kick
per atom necessarily involves a characterisation of the number of atoms
in the set, given that only the total momentum is known. Since the atom number
will not be measured, a priori information is required for estimating the
actual atom number. Clearly, this way atom-number fluctuations, whether
classical or quantum in nature, cannot be coped with, which shows an intrinsic
source of imperfection of the method.

Furthermore, since a measurement on a single system and not a series of
measurements on identically prepared systems is performed, not ensemble
averages are measured. Depending on the measurement resolution strong
correlations between the atoms are induced by the measurement projection,
since a huge number of microstates of the atoms may be associated to the same
observed measurement outcome, which form a complicated superposition state.

\subsection{Single-atom density matrix}
\label{sec:density}

\begin{figure}
  \begin{center}
    \includegraphics[width=0.5\textwidth]{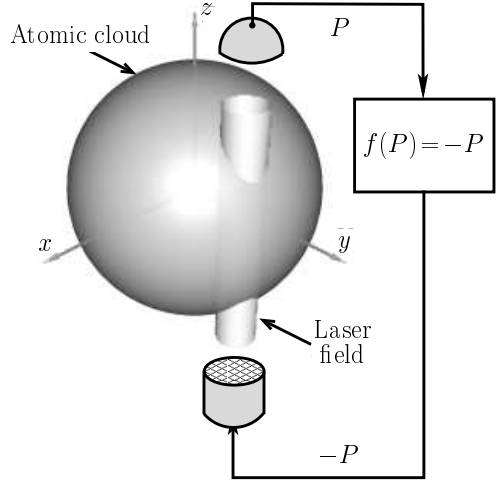}    
    \caption{Geometry of the feedback setup. The laser beam is aligned
      along the $z$ axis, it determines the size of the feedback
      region.}
    \label{fig:geometry}
  \end{center}
\end{figure}

Here we consider a full three-dimensional model and consider spatial
confinement of the volume where atoms are manipulated, see
Fig.~\ref{fig:geometry}. For simplicity we assume only one laser-beam profile,
despite the fact that several laser beams are involved in the implementation
of the required operations
\cite{recoil-induced,recoil-induced2,recoil-induced3}. The control-laser beam
is directed along the $z$-axis and its transverse profile in $x$ and $y$
directions is described by the beam-waist function $w_\perp(\vec{r}) \!=\!
w_\perp(x,y)$. Thus, the $z$ component of the total momentum of atoms inside
the beam $\hat{P}_w$ is measured and then compensated to zero by means of a
negative feedback loop. Using the atomic field operator $\hat{\phi} (\vec{r})$
for bosonic atoms, i.e. with commutator
\begin{equation}
\label{eq:boson-field}
[\hat{\phi}
(\vec{r}), \hat{\phi}^\dagger (\vec{r}')] \!=\! \delta(\vec{r} \!-\!
\vec{r}') , 
\end{equation}
this observable can be written as ($\hbar \!=\! 1$)\footnote{Throughout the
  paper we use $\hbar = 1$.}
\begin{equation}
  \label{eq:momentum}
  \hat{P}_w = -i \int \! dV \, w_\perp(\vec{r}) \, \hat{\phi}^\dagger(\vec{r})
  \, \partial_z \hat{\phi}(\vec{r}) .
\end{equation}

The many-body quantum state of the atomic cloud after a single operation of
stochastic cooling can be given as an integral over all possible measurement
outcomes $P$ for the measured total momentum:
\begin{equation}
  \label{eq:uncond_rho}
  \hat{\varrho}_+ = \int \! dP \, \hat{U}(P) \, \hat{M}(P) \, \hat{\varrho}_-
  \, \hat{M}^\dagger(P) \, \hat{U}^\dagger(P) .
\end{equation}
In this expression the initial many-body density operator is denoted as
$\hat{\varrho}_-$ and the final one, after the single feedback operation, is
denoted as $\hat{\varrho}_+$. It is assumed here, that the measurement and
shift of momentum can be performed on a time scale $\Delta t$ much faster than
the characteristic dynamics of the free system, i.e. $\Delta t \!\ll\!
\omega^{-1}$ with $\omega$ being the trap frequency. Then measurement and
shift can be taken as instantaneous processes without time delay between them.

The measurement of momentum $\hat{P}_w$ with specific outcome $P$ is 
described by the resolution amplitude \cite{res-amplitude,res-amplitude2}
\begin{equation}
  \label{eq:POVM}
  \hat{M}(P) = \sqrt{\frac{1}{\sqrt{2 \pi} \sigma}} \, {\rm exp}
  \left\{ -\frac{(P \!-\! \hat{P}_w )^2}{4 \, \sigma^2}\right\} ,
\end{equation}
where $\sigma$ denotes the measurement resolution. Applied on a momentum
eigenstate $| P_0 \rangle$ it gives the probability amplitude to observe the
value $P$.\footnote{The set of operators $\hat{M}^\dagger(P) \hat{M} (P)$
  forms a positive operator-valued measure \cite{neumark}.} A quasi
canonically conjugate operator for the centre of mass of the atoms in the
laser beam, being experimentally accessible, can be defined analogously as
\cite{wal-feedback}
\begin{equation}
  \label{eq:coordinate}
  \hat{Q}_w = \frac{1}{N_e} \int \! dV \, w_\perp(\vec{r}) \,
  \hat{\phi}^\dagger(\vec{r}) \, z \, \hat{\phi}(\vec{r}) .
\end{equation}
Since $N_e$ is an estimated atom number, the commutator relation of operator
$\hat{P}_w$ and $\hat{Q}_w$ reveals a deviation from the usual canonical form:
\begin{equation}
  \label{eq:commutator}
  [ \hat{Q}_w , \hat{P}_w ] = i \hat{N}_w / N_e .
\end{equation}
Here the true atom-number is defined as the operator
\begin{equation}
  \label{eq:number}
  \hat{N}_w = \int \! dV \, w_\perp^2(\vec{r}) \, \hat{\phi}^\dagger(\vec{r})
  \hat{\phi}(\vec{r}) .
\end{equation}
The modified commutation relation (\ref{eq:commutator}) has an impact on the action of the shift operator
$\hat{U}(P)$, that is supposed to produce a shift of $\hat{P}_w$ by $-P$, and
that is defined as 
\begin{equation}
  \label{eq:shift}
  \hat{U}(P) = \exp \!\big( -i P \hat{Q}_w \big) .
\end{equation}
Transforming the observable $\hat{P}_w$ by use of to the
unitary transformation~(\ref{eq:shift}) we obtain
\begin{equation}
  \label{eq:P-shift}
  \hat{U}^\dagger(P) \, \hat{P}_w \, \hat{U}(P) = \hat{P}_w - P \hat{N}_w /
  N_e .  
\end{equation}
Thus an optimal shift by $-P$ is produced only on average when additionally
estimating 
\begin{equation}
\label{eq:opt-estimate}
N_e \!=\! \langle \hat{N}_w \rangle .
\end{equation}
Nevertheless, atom-number fluctuations will always deteriorate the perfection
of the shift operation. In view of the lack of knowledge on the true atom
number $\hat{N}_w$, the estimate (\ref{eq:opt-estimate}) represents an
optimum. Thus in the following we use this justifiable estimate.

For our further derivations it is convenient to calculate the single-atom
density matrix from Eq.~(\ref{eq:uncond_rho}). It is defined as
\begin{equation}
  \label{eq:single-atom-dm}
  \sigma(\vec{r}_1, \vec{r}_2) = \langle \hat{\phi}^\dagger(\vec{r}_2)
  \hat{\phi}(\vec{r}_1) \rangle ,
\end{equation}
and can be calculated as a trace over the many-body density operator given in
Eq.~(\ref{eq:uncond_rho}). In this way the single-atom density matrix after
$(+)$ an operation of stochastic cooling reads
\begin{equation}
  \label{eq:reduced}
  \sigma_+(\vec{r}_1, \vec{r}_2) = \int \! dP \left\langle 
  \hat{M}^\dagger(P) \, \hat{U}^\dagger(P) \,
  \hat{\phi}^\dagger(\vec{r}_2) \, \hat{\phi}(\vec{r}_1)
  \, \hat{U}(P) \, \hat{M}(P) \right\rangle_- ,
\end{equation}
where $\langle \ldots\rangle_-$ denotes tracing over the many-body density
operator $\hat{\varrho}_-$, that represents the quantum state before the
feedback operation. 

The action of $\hat{U}(P)$ on a field operator results as a c-number
exponential factor
\begin{equation}
  \label{eq:U-transform}
  \hat{U}^\dagger(P) \, \hat{\phi}(\vec{r}) \, \hat{U}(P) = 
  \hat{\phi}(\vec{r}) \, \exp \!\left[ - i z  w_\perp(\vec{r}) P / \langle
  \hat{N}_w  \rangle  \right] . 
\end{equation}
Moreover, using the Fourier representation of the resolution amplitude 
\begin{equation}
  \label{eq:M-fourier}
  \hat{M}(P) = \int \! dq \, \underline{M}(q) \, e^{i q (P -
  \hat{P}_w )}  , 
\end{equation}
with 
\begin{equation}
  \label{eq:M-fourier-def}
  \underline{M}(q) =  \sqrt[4]{\frac{2 \sigma^2}{\pi}} \exp(-\sigma^2
  q^2) 
\end{equation}
the single-atom density matrix can be rewritten as
\begin{eqnarray}
  \label{eq:reduced2}
  \fl 
  \sigma_+(\vec{r}_1, \vec{r}_2) & = & \int \! dP \! \int \! dq \!\int \! dq' 
  \left\langle \  e^{i q \hat{P}_w} \hat{\phi}^\dagger(\vec{r}_2)
    \hat{\phi}(\vec{r}_1) \, e^{-i q' \hat{P}_w} \right\rangle_- 
  \underline{M}^\ast(q) \, \underline{M}(q') 
  \nonumber \\
  \fl & & \times \, \exp \! \left\{ i P \left[ \frac{z_2 w_\perp(\vec{r}_2) 
  \!-\! z_1 w_\perp(\vec{r}_1)}{\langle \hat{N}_w \rangle} + q' \!-\! q
  \right] \right\} .
\end{eqnarray}
This result can be further simplified using the transformation
\begin{equation}
  \label{eq:P_action}
  \exp \left[ i q \hat{P}_w \right] \, \hat{\phi}(\vec{r}) \, \exp \left[ -i q
  \hat{P}_w \right] 
  = \hat{\phi}\big( x, y, z \!-\! q w_\perp(\vec{r}) \big) ,
\end{equation}
which results in
\begin{eqnarray}
  \label{eq:reduced3}
  \fl \sigma_+(\vec{r}_1, \vec{r}_2) = \int \! dP \int \! dq \int \! dq' \,
  \underline{M}^\ast(q) \underline{M}(q') 
  \exp \! \left\{ i
  P \left[ \frac{z_2 w_\perp(\vec{r}_2) \!-\! z_1 w_\perp(\vec{r}_1)}{\langle
  \hat{N}_w \rangle} +
  q' \!-\! q \right] \right\} \nonumber \\
\times \, \left\langle \hat{\phi}^\dagger \big(x_2, y_2, z_2 \!-\! q
  w_\perp(\vec{r}_2)\big) \, \hat{\phi} \big(x_1, y_1, z_1 \!-\! q
  w_\perp(\vec{r}_1) \big) \,
    e^{i (q \!-\! q') \hat{P}_w} \right\rangle . 
\end{eqnarray}
Performing then the $P$ and $q'$ integrations we finally obtain
\begin{eqnarray}
  \label{eq:single-atom-density-matrix}
  \fl \sigma_+(\vec{r}_1, \vec{r}_2) = 2 \pi \! \int \! dq \,
  \underline{M}^\ast(q) \underline{M}\big(q \!+\! [z_1 w_\perp(\vec{r}_1)
  \!-\! z_2 w_\perp(\vec{r}_2)]/ \langle \hat{N}_w \rangle \big) \nonumber \\
  \fl \times \left\langle
  \hat{\phi}^\dagger \big( x_2, y_2, z_2 \!-\! q w_\perp(\vec{r}_2) \big)
  \, \hat{\phi} \big(x_1, y_1, z_1 \!-\! q w_\perp(\vec{r}_1) \big) 
  \, e^{i \hat{P}_w 
  [z_2 w_\perp(\vec{r}_2) - z_1 w_\perp(\vec{r}_1)] / \langle \hat{N}_w
  \rangle} \right\rangle_- ,
\end{eqnarray}
which shows that the single-atom density matrix after the feedback depends in
general on higher atom-atom correlations before the feedback. In
Eq.~(\ref{eq:single-atom-density-matrix}) this is encoded by the occurrence of
the exponential operator.

\section{Feedback-induced change of energy}
\label{sec:energy-change}

Important features of the application of a single feedback step can be given
in analytical form. For example, the difference of energy after and before the
feedback step $\Delta E$ can be calculated from
Eq.~(\ref{eq:single-atom-density-matrix}). The information contained therein
allows us to recognise noise sources and determine optimal parameters for
maximum cooling. The parameters that can be optimised are the measurement
resolution $\sigma$ and the geometrical characteristics given by the size and
location of the control-laser beam with respect to the trapping potential.

In the following we use the thermal equilibrium state of the atomic ensemble
to calculate the average energy change. This allows us to obtain a natural
description of cooling in terms of energy $\Delta E (T)$ that is subtracted
or, possibly, added by a feedback operation at a certain temperature point.
However, this approach does not necessarily reflect the most general
experimental situation, since specific correlations generated step by step in
the feedback process are not taken into account. These correlations may
possibly lead to
enhancement of cooling via various effects, as shown in
Refs~\cite{stochastic-raizen,stochastic-raizen-josa}. Thus the results of
this paper, where we assume a thermal equilibrium state, represent the leading
cooling/heating mechanisms for the quasi-equilibrium case. 

Having the expression for the single-atom density matrix
(\ref{eq:single-atom-density-matrix}) the change of total energy due to the
application of a single feedback step can be formulated. The Hamiltonian of
the system of non-interacting atoms in the isotropic, harmonic trap potential
is
\begin{equation}
  \label{eq:hamilton}
  \hat{H} = \int \! dV \, \hat{\phi}^\dagger(\vec{r}) 
  \left[ - \frac{\nabla^2}{2m} 
    + \frac{m\omega^2}{2} \, \vec{r}^2 \right] 
  \hat{\phi}(\vec{r}) ,
\end{equation}
where $m$ and $\omega$ are the atomic mass and the vibrational trap frequency,
respectively. We divide this Hamiltonian into parts
describing the energy of the motion in $z$ direction, i.e. in longitudinal
direction with respect to the measured momentum $\hat{P}_w$, and into
parts related to the energy of the transverse motion in the $xy$ plane.

The average total energy of the system in a given many-body quantum
state can be written as the sum of these different contributions as
\begin{equation}
  \label{eq:E-def}
  E = \langle \hat{H} \rangle = T_\parallel + T_\perp + V_\parallel + V_\perp
  . 
\end{equation}
Using the definition of the single-atom density
matrix~(\ref{eq:single-atom-dm}), the corresponding kinetic parts of
Eq.~(\ref{eq:E-def}) are given as
\begin{eqnarray}
  \label{eq:K-parallel-def}
  T_\parallel & = & - \frac{1}{2m} \int \! dV \! \int \! dV' \, \delta(\vec{r} 
  \!-\! \vec{r}') \, \partial_z^2 \, \sigma(\vec{r}, \vec{r}') , \\
  \label{eq:K-perp-def}
  T_\perp & = & - \frac{1}{2m} \int \! dV \, \delta(\vec{r} \!-\! \vec{r}') \, 
  \nabla_\perp^2
  \, \sigma(\vec{r}, \vec{r}') ,
\end{eqnarray}
where $\nabla_\perp^2 \!=\! \partial_x^2 \!+\! \partial_y^2$, and the
potential-energy contributions read
\begin{eqnarray}
  \label{eq:V-parallel-def}
  V_\parallel & = & \frac{m\omega^2}{2} \int \! dV \, z^2 \,
  \sigma(\vec{r}, \vec{r})  , \\
  \label{eq:V-perp-def}
  V_\perp & = & \frac{m\omega^2}{2} \int \! dV \, (x^2 \!+\! y^2) \,
  \sigma(\vec{r}, \vec{r})  .
\end{eqnarray}

A change of the average energy due to the application of a single step of
stochastic cooling is dominantly generated by the energy exchange with the
externally applied optical fields that implement the momentum shift of atoms.
The unidirectional flow of energy from the system to the optical fields is
determined by the irreversibility introduced in the quantum measurement
process.  Apart from that, however, there are several sources of heating, among
them also the back-action noise of the measurement itself. Especially at
ultralow temperatures a decay or even reversal of the net energy flow may be
expected, when the detrimental heating terms compensate the sought cooling
effect of the momentum shift. In the following we will extract all these
energetic terms by considering the feedback-induced change of energy based on
Eqs~(\ref{eq:E-def})--(\ref{eq:V-perp-def}). More specifically, we consider the
change of the average energy, i.e. the difference between the energy after a
single step of stochastic cooling and that before,
\begin{equation}
  \label{eq:DE-total}
  \Delta E = E_+ \!-\! E_- = \Delta T_\parallel + \Delta T_\perp + \Delta
  V_\parallel + \Delta V_\perp .
\end{equation}

\subsection{Energy change in the longitudinal motion}

The dominant change of energy will occur in the potential and kinetic energies
associated with the longitudinal motion in $z$ direction. The longitudinal
potential energy $V_\parallel$ is given by Eq.~(\ref{eq:V-parallel-def}) and
together with Eq.~(\ref{eq:single-atom-density-matrix}) it can be shown that
the change of longitudinal potential energy is
\begin{equation}
  \label{eq:DV-parallel}
  \Delta V_\parallel = \frac{m\omega^2}{8 \sigma^2} \langle \hat{N}_w \rangle
  . 
\end{equation}
This positive energy contribution arises from the back-action noise of the
total-momentum measurement in $z$ direction. The centre-of-mass $z$ coordinate
of the affected atoms, associated with the total mass $m \langle \hat{N}_w
\rangle$, is then subject to an increased uncertainty of the size
$(2\sigma)^{-1}$, which Eq.~(\ref{eq:DV-parallel}) shows to introduce a
heating term in the potential energy.

For the kinetic energy of the motion in $z$ direction a more involved
calculation of the second-order $z$-derivative of the density
matrix~(\ref{eq:single-atom-density-matrix}) is required.  After some lengthy
but straightforward calculation the following expression for the
feedback-induced change of longitudinal kinetic energy is then obtained:
\begin{equation}
  \label{eq:DK-parallel}
  \Delta T_\parallel = \frac{\sigma^2}{2 m \langle
  \hat{N}_w \rangle}  -
  \frac{\langle \hat{P}_w^2 \rangle}{2m \langle \hat{N}_w \rangle} +
  \frac{\langle \Delta \hat{N}_w  \hat{P}_w^2
  \rangle}{2m \langle \hat{N}_w \rangle^2} .
\end{equation}
Here $\Delta \hat{N}_w \!=\! \hat{N}_w \!-\! \langle \hat{N}_w \rangle$ is the
fluctuation of the actual atom number in the control beam around its average.
The first term in Eq.~(\ref{eq:DK-parallel}) is the kinetic energy left in the
system by the imprecise total-momentum measurement with resolution $\sigma$.
The second term is the sought cooling effect, where the centre-of-mass kinetic
energy of the affected atoms is removed from the system. The last term,
though, arises from quantum fluctuations of the number of atoms in the
control-laser beam.  This heating term appears since in the momentum-shift
operation the actual atom number $\hat{N}_w$ is not known but only estimated
by $\langle \hat{N}_w \rangle$. Thus this term represents a
quantum-statistical imperfection of the feedback loop of stochastic cooling.

\subsection{Energy change in the transverse motion}

At first sight one may guess that the transverse motion is not affected by the
feedback loop, since only momentum in $z$ direction is measured and shifted.
However, since the atoms that contribute to the measured signal are confined
within the laser beam waist $w_\perp(\vec{r})$, the measurement of $\hat{P}_w$
also contains an indirect measurement of the transverse position of atoms with
a resolution roughly given by the diameter of the beam. Thus a back-action
noise in the transverse momenta can be expected that may lead to further
contributions to the kinetic energy.  It is now left to show how large these
energy contributions are compared with those emerging from the longitudinal
motion. 

From Eqs~(\ref{eq:single-atom-density-matrix}) and (\ref{eq:V-perp-def}) it
can be easily seen that the potential energy in transverse $x$ and $y$
directions is unchanged, i.e.
\begin{equation}
  \label{eq:DV-perp}
  \Delta V_\perp = 0 .
\end{equation}
This result is obvious since only the momentum in $z$ direction with a
transverse spatial confinement is measured without affecting the noise in the
transverse coordinates. Let us therefore consider the kinetic energy of the
transverse coordinates as defined in Eq.~(\ref{eq:K-perp-def}).  Calculating
the required second-order derivatives of
Eq.~(\ref{eq:single-atom-density-matrix}) and performing the integrations,
after some lengthy but straightforward calculus, we obtain for the change in
transverse kinetic energy
\begin{eqnarray}
  \label{eq:K-perp}
  \fl \Delta T_\perp & = & \frac{1}{2 m} \int \! dV \, [\nabla
  w_\perp(\vec{r})]^2  \bigg\{ \frac{m}{2\sigma^2} \langle
  \hat{T}_\parallel(\vec{r}) \rangle + \frac{1}{\langle \hat{N}_w \rangle^2}
  \left[\sigma^2 z^2 \!+\! {\textstyle\frac{3}{4}} \, 
    w_\perp^2(\vec{r})  \right] \langle \hat{\phi}^\dagger(\vec{r})
  \hat{\phi}(\vec{r}) \rangle  \nonumber \\
  \fl & & + \frac{1}{2 \sigma^2 \langle \hat{N}_w \rangle^2} \left[ \sigma^2
  z^2 + {\textstyle\frac{1}{4}} \, w_\perp^2(\vec{r}) \right]  \langle \{
  \hat{\phi}^\dagger(\vec{r}) \hat{\phi}(\vec{r}), \hat{P}_w^2 \} \rangle 
  \\
  \fl & & + \frac{1}{4 \sigma^2 \langle \hat{N}_w \rangle}
  w_\perp(\vec{r}) \langle \{ \hat{p}_z(\vec{r}),
  \hat{P}_w \} \rangle \bigg\}
  - \frac{1}{2 m \langle \hat{N}_w \rangle} \int \! dV \, z \, \nabla
  w_\perp(\vec{r}) \!\cdot\!  \langle \{ \hat{\vec{p}}(\vec{r}) , \hat{P}_w \}
  \rangle , \nonumber 
\end{eqnarray}
where $\{ \hat{A} , \hat{B} \} \!=\! \hat{A} \hat{B} \!+\! \hat{B} \hat{A}$ is
the anti-commutator and the momentum density reads
\begin{equation}
  \label{eq:vector-p}
  \hat{\vec{p}}(\vec{r}) = - \frac{i}{2} \left\{ \hat{\phi}^\dagger (\vec{r})
  \nabla \hat{\phi} (\vec{r}) - \left[\nabla \hat{\phi}^\dagger (\vec{r})
  \right]  \hat{\phi} (\vec{r}) \right\} .
\end{equation}
For a thermal equilibrium state, space dependent averages
will have a symmetry with respect to $z \!\to\! -z$, and thus the second
integral in Eq.~(\ref{eq:K-perp}) can be shown to vanish as an odd moment of
$z$. 

\subsection{Gaussian beam-waist function}
At this point the specific form of the beam-waist function shall be
introduced. We consider here a Gaussian beam with the following definition for
$w_\perp(\vec{r})$:
\begin{equation}
  \label{eq:w-perp-def}
  w_\perp(\vec{r}) = \exp\!\left[- \frac{(x \!-\! x_0)^2 \!+\! (y \!-\!
  y_0)^2}{r_0^2}  \right] .
\end{equation}
In this way the area $A_0$ of the integrated beam intensity,
\begin{equation}
  \label{eq:area}
  A_0 = \int \! dA \, w_\perp^2(\vec{r}) = \pi r_0^2 ,
\end{equation}
allows us to interpret $r_0$ as an effective radius of the laser beam. The
squared gradient of $w_\perp(\vec{r})$ results from Eq.~(\ref{eq:w-perp-def})
as
\begin{equation}
  \label{eq:sqr-grad}
  [\nabla w_\perp(\vec{r})]^2 = w_\perp^2(\vec{r}) \, \frac{
  (x \!-\! x_0)^2 \!+\! (y \!-\! y_0)^2}{r_0^4} ,
\end{equation}
and using these results, Eq.~(\ref{eq:K-perp}) reads
\begin{eqnarray}
  \label{eq:K-perp2}
  \fl \Delta T_\perp & = & \frac{1}{2 m} \int \! dV \,
  w_\perp^2(\vec{r}) \bigg\{
  \frac{m}{2\sigma^2} \langle
  \hat{T}_\parallel(\vec{r}) \rangle 
  + \frac{1}{\langle \hat{N}_w \rangle^2} \left[\sigma^2 z^2 \!+\!
    {\textstyle\frac{3}{4}} \, 
    w_\perp^2(\vec{r})  \right] 
  \langle \hat{\phi}^\dagger(\vec{r}) \hat{\phi}(\vec{r}) \rangle 
  \nonumber \\
  \fl & & + \frac{1}{2 \sigma^2 \langle \hat{N}_w \rangle^2} \left[ \sigma^2
  z^2  + {\textstyle\frac{1}{4}} \, w_\perp^2(\vec{r}) \right] \langle \{
  \hat{\phi}^\dagger(\vec{r}) \hat{\phi}(\vec{r}), \hat{P}_w^2 \} \rangle 
  \nonumber \\
  \fl & & + \frac{1}{4 \sigma^2 \langle \hat{N}_w \rangle}
  w_\perp(\vec{r}) \langle \{ \hat{p}_z(\vec{r}), \hat{P}_w \}
  \rangle   \bigg\}  \, \frac{
  (x \!-\! x_0)^2 \!+\! (y \!-\! y_0)^2}{r_0^4}  .
\end{eqnarray}
This kinetic-energy change will determine the heating effect due to the
transverse confinement of atoms in the laser beam. 

\section{Non-degenerate atomic vapour}
\label{sec:non-degenerate-atomic-vapor}

For ultracold temperatures near the condensation temperature $T_0$ the
longitudinal energy change has been discussed already in the approximation of
a rectangular shape of the beam waist in Ref.~\cite{stochastic}. In the
following we evaluate the complete energy change (\ref{eq:DE-total}), with the
contributions given by Eqs.~(\ref{eq:DV-parallel}), (\ref{eq:DK-parallel}),
(\ref{eq:DV-perp}) and (\ref{eq:K-perp2}). We consider the regime of a
non-degenerate gas, where the thermal de Broglie wavelength is much smaller
than the interatomic distance.  In that case the feature of
indistinguishability of atoms can be neglected, keeping however the full
wavemechanics of the single atom. That is, the single atom's position and
momentum still obey the canonical commutator relation, from which several
important effects emerge.

Calculating expectation values for a thermal state in
the canonical ensemble at temperature $T$ and total atom number
$N$, treating the atoms as distinguishable particles, the number of
atoms in the laser beam, for example, results as
\begin{equation}
  \label{eq:Nw}
  \langle \hat{N}_w \rangle = N \frac{ s^2}{2
  \!+\! s^2} \exp\!\left( - \frac{d^2}{2 \!+\! s^2} \right) .
\end{equation}
Here we used the scaled distance $d$ of the control beam from the trap origin
and the
scaled beam radius $s$, defined by
\begin{equation}
  \label{eq:d-scaled}
  d = \sqrt{(x_0^2 \!+\! y_0^2)} / L_{\rm th}, \qquad s = r_0 / L_{\rm
  th} .
\end{equation}
The rms extension of the atomic cloud is given by
\begin{equation}
  \label{eq:Lth}
  L_{\rm th} = \Delta x_0 \left[ \tanh \left( \frac{\omega}{2 k_{\rm B} T}
  \right) \right]^{-1/2} , 
\end{equation}
with $\Delta x_0 \!=\! \sqrt{1 / (2 m \omega)}$ being the ground-state
position uncertainty in the trap potential and $k_{\rm B}$ being the Boltzmann
constant. In the following we also use the size of the atomic cloud in units of
the ground-state uncertainty:  
\begin{equation}
  \label{eq:lth}
l_{\rm th} = L_{\rm th} / \Delta x_0 .  
\end{equation}

\subsection{Longitudinal energy change}

The complete change of energy in the longitudinal motion in units
of vibrational energy quanta reads
\begin{eqnarray}
  \label{eq:dE-parallel}
  \Delta E_\parallel / \omega  & = & \frac{1}{4} \left[
      \frac{(\sigma / \Delta 
        p_0)^2}{\langle \hat{N}_w \rangle} + \frac{\langle \hat{N}_w
        \rangle}{(\sigma / \Delta p_0)^2} \right] - \frac{l_{\rm
        th}^2}{4} 
    \\ \nonumber 
    & & + \frac{l_{\rm th}^2}{4 N} \left\{ \frac{(2 \!+\!
        s^2)^2}{s^2 (4 \!+\! 
        s^2)} \exp\!\left[ \frac{4 d^2}{(2 \!+\! s^2) (4 \!+\! s^2)}
      \right] - 1 \right\} ,
\end{eqnarray}
where $\Delta p_0 \Delta x_0 =\frac{1}{2}$ so that $\Delta p_0 = \sqrt{m
  \omega / 2}$. The first term represents the measurement-induced noise leading
to an increase of kinetic and potential energies. Taking into account only the
energy change as given here this heating effect can be minimised by adapting the measurement resolution to the number
of atoms in the control-laser beam as
\begin{equation}
  \label{eq:sigma-opt}
  \sigma = \Delta p_0 \sqrt{ \langle \hat{N}_w \rangle} .
\end{equation}
The minimum heating due to this noise results then as one half energy quantum.

The sought cooling effect is represented by the second term in
Eq.~(\ref{eq:dE-parallel}), which is the centre-of-mass kinetic energy of the
atoms addressed by the feedback. This value does neither depend on size nor
location of the feedback region. For large temperatures $l_{\rm th} \!
\rightarrow \! \sqrt{2 k_{\rm B} T / \omega}$, so that the subtracted kinetic
energy reduces to $k_{\rm B} T / (2 \omega)$, manifesting the removed energy
as being given by the equipartition theorem.

\begin{figure}
  \begin{center}
    \includegraphics[width=0.5\textwidth]{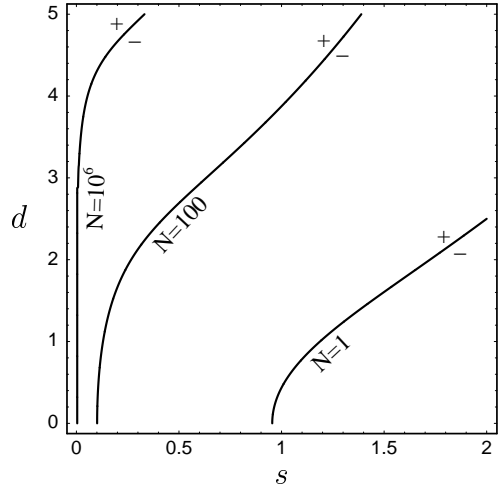} 
  \end{center}
  \caption{Boundaries between cooling (-) and heating (+) for the
    temperature $T \!=\! 10 \, T_0$ and varying total numbers of atoms.
    $\sigma$ is chosen as the optimal value given in
    Eq.~(\ref{eq:sigma-opt}).}
  \label{fig:long}
\end{figure}

Finally, the third term in Eq.~(\ref{eq:dE-parallel}) represents quantum noise
due to the transverse spatial confinement of atoms subject to the feedback. It
is produced by atom-number fluctuations that crucially depend via the average
atom number on the size and location of the control-laser beam. This noise is
always positive, leading thus to an unavoidable heating contribution, and does
vanish only for $s \!\to\! \infty$.  The limit $s \!\to\! \infty$ realizes the
situation where all atoms are inside the control-laser beam, thus containing
exactly $N$ atoms with vanishing atom-number fluctuations. Moreover, the
strength of this heating term diminishes with increasing total number of
atoms, as can be observed in Fig.~\ref{fig:long}.  

There the contours $\Delta E_\parallel \!=\! 0$ have been plotted in the
parameter space $(s,d)$ for varying total atom numbers at fixed temperature $T
\!=\!  10 \, T_0$, where
\begin{equation}
  \label{eq:T0}
  T_0 = \frac{\omega}{k_{\rm B}} \left( \frac{N}{\zeta (3)} \right)^{1/3},
\end{equation}
is the condensation temperature in the thermodynamic limit \cite{trapped_bec}
with $\zeta (n)$ being the Riemann $\zeta$ function. These contours represent
the boundaries between cooling (-) on the right-hand side of the contour and
heating (+) on its left-hand side.  They show that finer spatial resolutions
$s$ and larger distances from the trap centre $d$ are allowed for cooling when
the total atom number increases.
Whereas the measurement-induced noise is constant and negligibly
small, the major contribution leading to a restriction of the
parameters $s$ and $d$ comes from the atom-number fluctuations in the
chosen beam waist. 

\subsection{Transverse energy change}

The energy change in the transverse motion can be obtained in the same way as
that for the longitudinal one, cf. Eq.~(\ref{eq:dE-parallel}), and reads
\begin{eqnarray}
  \label{eq:dE-transverse}
  \fl  \Delta E_\perp / \omega & = & \frac{N}{4 \langle \hat{N}_w
        \rangle} \left[ \frac{\langle \hat{N}_w \rangle}{ 
        (\sigma / \Delta p_0)^2} + \frac{(\sigma / \Delta p_0)^2}{ \langle
        \hat{N}_w \rangle}
    \right]\frac{4 + s^2 (2 +
      d^2)}{(2 + s^2)^3} \exp\!\left[ - \frac{d^2}{2 + s^2} \right]
  \nonumber \\
  \fl & & + \frac{N}{4 \langle \hat{N}_w \rangle} \left[ \frac{l_{\rm th}^2 +
      l_{\rm th}^{-2}}{\langle \hat{N}_w \rangle} + \frac{2}{(\sigma / \Delta
      p_0)^2} \right] \frac{8 + s^2 (2 + d^2)}{(4 + s^2)^3} \exp\!\left[ -
    \frac{2 d^2}{4 + s^2} \right] 
  \nonumber \\
  \fl & &  + \frac{N ( N \!-\! 1)}{\langle \hat{N}_w \rangle^2} \frac{l_{\rm
        th}^2}{4}  
  \frac{s^2 [ 4 + s^2 (2 + d^2) ]}{(2 +
    s^2)^4} \exp\!\left[ -
    \frac{2d^2}{2 + s^2} \right] 
  \nonumber \\
  \fl & & +  
  \frac{1}{4 (\sigma/\Delta p_0)^2 \langle \hat{N}_w \rangle^2} \Bigg\{ 
  N \frac{12 + s^2 (2 + d^2)}{(6 + s^2)^3} \exp\!\left[ -
    \frac{3 d^2}{6 + s^2} \right] 
  \nonumber \\
  \fl & & + N ( N \!-\! 1) \frac{s^2}{2 + s^2} \frac{8 + s^2 (2+
    d^2)}{(4 + s^2)^3} \exp\!\left[ -
    \frac{d^2 (8 + 3 s^2)}{(4 + s^2) ( 2 + s^2)} \right] \Bigg\} .
\end{eqnarray}
As mentioned before these heating contributions vanish in the limit $s
\rightarrow \infty$. Moreover, they depend on the total number of atoms $N$,
which is also mediated by the dependence on $\langle \hat{N}_w \rangle$, see
Eq.~(\ref{eq:Nw}), and possibly on $\sigma$.
  
\begin{figure}
  \begin{center}
    \includegraphics[width=0.5\textwidth]{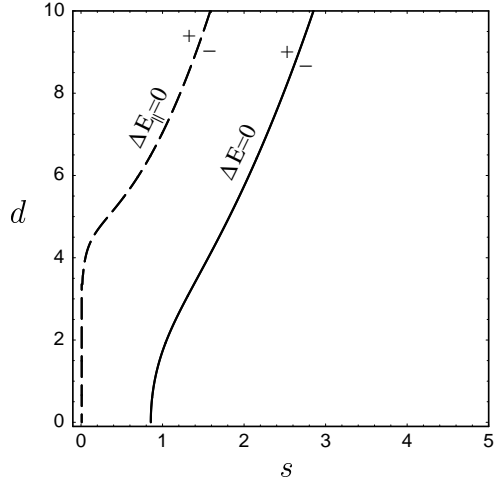} 
  \end{center}
  \caption{Contours $\Delta E_\parallel \! =\! 0$ (dashed lines) and $\Delta
    E \! = \! 0$ (solid lines) for different temperatures and $ N \!
    = \!  10^6 $.}
  \label{fig:trans1}
\end{figure}

Let us first consider the emerging changes in the boundary between cooling and
heating, when now also the transverse energy change is taken into account.
That is, we look for the contour in $(s,d)$ parameter space that satisfies
$\Delta E \!=\! 0$, where $\Delta E \!=\! \Delta E_\parallel \!+\! \Delta
E_\perp$. In Fig.~\ref{fig:trans1} this contour (solid curve) is shown and
compared to the boundary based only on longitudinal terms (dashed curve) for a
total atom number of $N \!=\! 10^6$.
It is clearly observed that the additional heating terms due to the
transverse confinement of atoms in the control beam leads to a shift
of the boundary to larger values of $s$ and smaller values of $d$. The
former is due to the fact that the measurement of the total
momentum in $z$ direction of atoms in the beam, indirectly also
represents a measurement of transverse coordinates within the beam-waist
size. This leads again to measurement back-action noise, which is naturally
reduced by increasing $s$. 

The fact that the boundary is shifted to smaller values of $d$ is due to
relative atom-number fluctuations that decrease for increasing atom numbers
found near to the trap centre. It should be noted that either curve actually
contains two different curves at temperatures $T \!=\! 10 \, T_0$ and $T \!=\!
10000 \, T_0$, which however cannot be distinguished in the plot. The only
explicit dependence in Eqs.~(\ref{eq:dE-parallel}) and
(\ref{eq:dE-transverse}) is due to the occurrence of $l_{\rm th}$. For $N \!
\gg \! 1$, however, this dependence is very weak. Nevertheless, all features
discussed here and in the following implicitly depend on temperature via the
chosen scaling of $s$ and $d$ by $L_{\rm th}$.

\begin{figure}
  \begin{center}
    \includegraphics[width=0.5\textwidth]{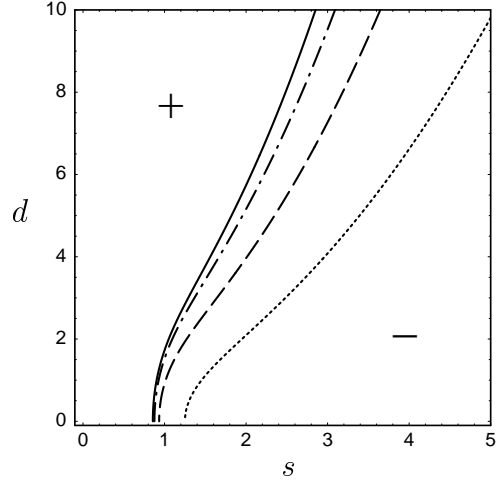}
  \end{center}
  \caption{Boundaries $\Delta E \! = \! 0$ for $T  \! = \! 10 \, T_0$ and 
    varying total atom numbers: $N \!=\! 1$ (dotted curve), $N \!=\! 10$
    (dashed curve), $N \!=\! 100$ (dot-dashed curve), $N \!=\! 10^6$ (solid
    curve).}
  \label{fig:trans2}
\end{figure}

Assuming that for increasing total atom number the optimised measurement
resolution $\sigma$ increases, as for the value given in
Eq.~(\ref{eq:sigma-opt}) for longitudinal terms, for large atom numbers $N
\!\gg\! 1$ an asymptotic expansion can be given for
Eq.~(\ref{eq:dE-transverse}):
\begin{equation}
  \label{eq:dE-transverse-asymptotic}
   \Delta E_\perp / \omega = \frac{1}{4} \left[ \frac{\langle \hat{N}_w
        \rangle}{ (\sigma / \Delta p_0)^2} + \frac{(\sigma / \Delta p_0)^2}{
       \langle \hat{N}_w \rangle} + l_{\rm th}^2  \right]\frac{4 \!+\! s^2 (2
        \!+\! d^2)}{s^2 (2 \!+\! s^2)^2} , 
\end{equation}
where Eq.~(\ref{eq:Nw}) has been used.  In the same limit the longitudinal
energy change can be approximated to finally obtain the total change of energy
as
\begin{equation}
  \label{eq:dE-asymptotic}
\fl  \Delta E / \omega = 
  \frac{1}{4} \left[ \frac{(\sigma / \Delta 
      p_0)^2}{\langle \hat{N}_w \rangle} \!+\! \frac{\langle \hat{N}_w
      \rangle}{(\sigma / \Delta p_0)^2} \right] \left[ 1 \!+\! \frac{4
      \!+\! s^2 (2 \!+\! 
      d^2)}{s^2 (2 \!+\! s^2)^2} \right] 
  - \frac{l_{\rm
      th}^2}{4} \left[ 1 \!-\! \frac{4 \!+\! s^2 (2 \!+\!
      d^2)}{s^2 (2 \!+\! s^2)^2} \right] .
\end{equation}
This asymptotic expansion represents a good approximation starting already
from atom numbers $N \!>\! 100$, as can be seen from
Fig.~\ref{fig:trans2}. There it is observed that for fixed temperature the
boundaries converge quickly to the asymptotic one for $N \rightarrow \infty$.

\begin{figure}
  \begin{center}
    \includegraphics[width=0.5\textwidth]{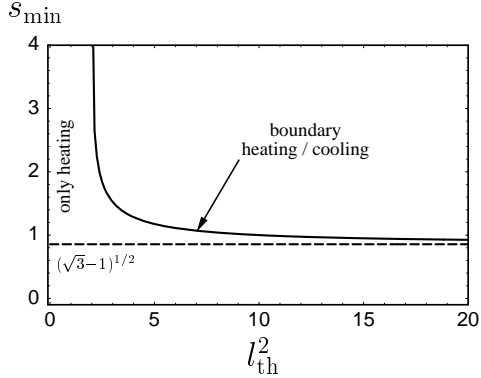}
  \end{center}
  \caption{Dependence of $s_{\rm min}$ on $l_{\rm th}^2$ for large total
    number of atoms $N \rightarrow \infty$.}
  \label{fig:temper}
\end{figure}

Equation~(\ref{eq:dE-asymptotic}) shows that again the optimal value
for the measurement resolution is given by (\ref{eq:sigma-opt}), for
which the final result reads
\begin{equation}
  \label{eq:dE-asymptotic-opt}
  \Delta E / \omega = 
  \frac{1}{2} \left[ 1 + \frac{4
      \!+\! s^2 (2 \!+\! 
      d^2)}{s^2 (2 \!+\! s^2)^2} \right] 
  - \frac{l_{\rm
      th}^2}{4} \left[ 1 - \frac{4 \!+\! s^2 (2 \!+\!
      d^2)}{s^2 (2 \!+\! s^2)^2} \right] .
\end{equation}
For this expression the solution of the condition $\Delta E \!=\! 0$ can be
analytically given as
\begin{equation}
  \label{eq:d(s)}
  d(s) = (2 \! + \! s^2) \sqrt{\frac{l_{\rm th}^2 \! - \! 2}{l_{\rm th}^2 \! +
  \! 2} - \frac{2}{s^2 (2 \!+\! s^2)}} .
\end{equation}
For $d \!=\! 0$ the corresponding value for $s$ is the minimal beam-waist
radius. This minimal radius is obtained as 
\begin{equation}
  \label{eq:s-min}
  s_{\rm min} = \left( \sqrt{1 \!+\! 2 \, \frac{l_{\rm th}^2 \!+\! 2}{l_{\rm
  th}^2 \!-\! 2}} - 1 \right)^{\frac{1}{2}} .
\end{equation}
In Fig.~\ref{fig:temper} this function is shown in dependence on $\l_{\rm
  th}^2$, which for high temperatures is proportional to $T$. At $l_{\rm th}^2
\!=\! 2$ the removed centre-of-mass kinetic energy exactly compensates the
measurement-induced heating, which requires $s_{\rm min} \rightarrow \infty$.
For values $\l_{\rm th}^2 \!<\! 2$ cooling does not occur, since the
unavoidable measurement-induced noise can no longer be compensated. However,
the corresponding temperatures for $l_{\rm th}^2 \!\leq\! 2$ are below the
condensation temperature $T_0$, where our approach for a non-degenerate gas is
no longer valid. For this regime see Ref.~\cite{stochastic}.

For larger values of $l_{\rm th}^2$, or $T$ correspondingly, the limiting value
$s_{\rm min} \rightarrow (\sqrt{3} \!-\! 1)^{1/2} \! \approx \! 0.86$ is
reached. In this regime only a fraction of the atomic cloud needs to be
subject to the feedback loop since $s_{\rm min} \!<\! 1$. The unscaled minimum
beam-waist radius is thus 86\% of the rms extension of the atomic cloud. 

\section{Conclusions}
\label{sec:conclusions}

In summary we have studied the effects of transverse confinement in stochastic
cooling of trapped atoms. It could be clearly shown that these effects are
substantial for the cooling process and that minimum values for both the size
and location of the control-laser beam exist.

In the regime of non-degenerated gases analytical expressions could be derived,
that contain the full quantum-fluctuation effects. Among these effects are
atom-number fluctuations that appear due to the finite volume of the control
beam. They appear in form of an imperfection of the feedback loop and in form
of back-action noise due to the indirect measurement of transverse coordinates.

\section*{Acknowledgements}

This research was supported by Deutsche Forschungsgemeinschaft.

\end{document}